# Charge relaxation dynamics of an electrolytic nanocapacitor

*Vaibhav Thakore*[†, ‡,] *and James J. Hickman*[†, ‡, §, *]

[†]Department of Physics, [‡]NanoScience Technology Center and [§]Department of Chemistry, University of Central Florida, 12424 Research Parkway, Suite 400, Orlando, FL 32826, United States.

*Corresponding author*. James J Hickman, Phone: 407-823-1925; Fax: 407-882-2819, E-mail: jhickman@ucf.edu

*Present address*: Vaibhav Thakore, COMP Centre of Excellence at the Department of Applied Physics, Aalto University, School of Science, FI-00076 Aalto, Espoo, Finland

ABSTRACT: Understanding ion relaxation dynamics in overlapping electric double layers (EDLs) is critical for the development of efficient nanotechnology based electrochemical energy storage, electrochemomechanical energy conversion and bioelectrochemical sensing devices as well as controlled synthesis of nanostructured materials. Here, a Lattice Boltzmann (LB) method is employed to simulate an electrolytic nanocapacitor subjected to a step potential at t = 0 for various degrees of EDL overlap, solvent viscosities, ratios of cation to anion diffusivity and electrode separations. The use of a novel, continuously varying and Galilean invariant, molecular speed dependent relaxation time (MSDRT) with the LB equation recovers a correct microscopic description of the molecular collision phenomena and enhances the stability of the LB algorithm. Results for large EDL overlaps indicated oscillatory behavior for the ionic current density in contrast to monotonic relaxation to equilibrium for low EDL overlaps. Further, at low solvent viscosities and large EDL overlaps, anomalous plasma-like spatial oscillations of the electric field were observed that appeared to be purely an effect of nanoscale confinement. Employing MSDRT in our simulations enabled a modeling of the fundamental physics of the transient charge relaxation dynamics in electrochemical systems operating away from equilibrium wherein Nernst-Einstein relation is known to be violated.

## INTRODUCTION

Electric double layers (EDLs) associated with biomolecules, polymers, charged surfaces and electrochemical interfaces play an important role in a wide variety of phase transformation[1-3], interface and transport phenomena[4-7] under application of time varying electric potentials or currents. Advancing an understanding of transient ion transport and relaxation dynamics in EDLs is therefore critical not just for the development of nanotechnology based efficient electrochemical energy storage[8-10], electrochemomechanical energy conversion[11], bioelectrochemical sensing[12,13], molecular trapping devices[14,15] but also for the synthesis of novel nanostructured materials through a control of EDL mediated phase transformations[2,3] and electrochemical reactions[16,17]. EDLs over the years have been modeled using equivalent circuit models[18,19], analytic solution and simulation of Poisson-Nernst-Planck system of equations[20-23] and more recently using ab initio molecular dynamics (MD) and Monte-Carlo (MC) methods[24-26]. While equivalent circuit models for the EDLs are clearly oversimplified and suffer from drawbacks[22,27,28], the analytical or numerical solutions of the Poisson-Nernst-Planck system of equations for describing the dynamics of the EDLs assume continuum dynamics[29,30]. As such, in mesoscale systems with overlapping EDLs and relatively large Knudsen numbers (ratios of molecular mean free path to the characteristic system length scale) the use of Poisson-Nernst-Planck approach becomes untenable[29,30]. However, MD and MC methods have been successful in predicting equilibrium EDL structure and non-equilibrium steady state properties of aqueous electrolytes[24-26], but, using MD and MC methods, the simulation of transient dynamics of charge relaxation in overlapping EDLs in response to applied electric potentials has remained a challenge. This is primarily because of the sheer computational expense required to track (a) individual ions and solvent molecules, and, (b) achieve statistical accuracy for electrolytic solutions where the ionic number densities are several orders of magnitude smaller than those for the solvent[25,31]. In this context, the use of a computationally efficient mesoscale simulation technique like Lattice Boltzmann Method (LBM), which is based on the Boltzmann transport equation and is valid at relatively large Knudsen numbers, holds promise[30].

**Lattice Boltzmann Equation and Nernst-Einstein relation.** In LBM, the time evolution of the single particle distribution function $f_{i,\alpha}$ of the $i^{th}$ specie of particles in an electrolyte mixture can be described using the non-dimensional discretized LBE[32-34] as

$$f_{i,\alpha}(\boldsymbol{r} + \boldsymbol{e}_\alpha \Delta t, t + \Delta t) = f_{i,\alpha}(\boldsymbol{r}, t) - \frac{f_{i,\alpha}(\boldsymbol{r}, t) - f_{i,\alpha}{}^{eq}(\boldsymbol{r}, t)}{\tau_i} + \Delta t F_i \quad (1)$$

where $t$ is the time, $\Delta t$ the time step, $\boldsymbol{r}$ the space coordinate, $\tau_i$ the relaxation time, $\boldsymbol{e}_\alpha$ the discrete particle velocity, $f_{i,\alpha}{}^{eq}$ the local equilibrium Maxwellian distribution, $F_i$ the discretized external force term and $i \rightarrow$ ($s$, $Cn$ and A$n$) corresponds to solvent particles, cations and anions respectively.

So far, however, application of LBM based modeling to transient simulations of EDLs in electrochemical systems and electroosmotic flows has not been attempted and assumptions of bulk electroneutrality and thin EDLs, or fully developed ionic concentration and electrostatic potential profiles for overlapping EDLs as initial conditions, have been employed to obtain non-equilibrium steady state solutions[4,35-38]. This is so, in part, because all these studies employ single relaxation times based on constant bulk ionic diffusion coefficients in the LBE for the simulation of ionic electrodiffusion. The use of a single relaxation time with the LBE in the Bhatnagar-Gross-Krook (BGK) approximation[39], corresponding to diffusion in the bulk electrolyte, implicitly assumes the validity of the Nernst-Einstein relation between the equilibrium ionic diffusion coefficients $D_i^{eq}$ and their mobilities $\mu_i^{eq}$. However, since the Nernst-Einstein relation ($D_i^{eq} = \mu_i^{eq} kT/z_i e$) is derived from purely equilibrium considerations, it is well established, both through experiments and simulations, that Nernst-Einstein relation is not valid for dynamics of charged particles in systems away from equilibrium[40-42]. As a result any attempt to simulate the transient dynamics of such highly nonlinear systems in the BGK approximation results in numerical instabilities. Thus, to account for deviations from the Nernst-Einstein relation that affect relaxation times for ionic distribution functions in the LBE and to simulate the transient dynamics of EDLs in electrochemical systems away from equilibrium, we propose here a molecular speed dependent relaxation time (MSDRT) based on Tait's theory of mean free path[43,44]. The proposed relaxation time, which is more generally applicable to hydrodynamic flow simulations as well, is first shown to recover a correct microscopic description of collision phenomena for the classic entrance flow problem of plane Poiseuille flow in a channel[45]. The MSDRT approach is then employed to simulate and understand the nature of transient charge relaxation dynamics in overlapping EDLs of an electrolytic nanocapacitor.

## METHODS

For all simulation results presented here, a two-dimensional nine-velocity D2Q9 model of LBM was employed. The simulations were carried out using message passing interface (MPICH 2.0) based parallel codes implemented in FORTRAN while post-processing and visualization of the results was carried out using MATLAB (The MathWorks, Natick, MA). The discretized equilibrium distribution functions used in the LBE were computed using

$$f_{i,\alpha}^{eq} = w_\alpha n_i \left[1 + \frac{\boldsymbol{e}_\alpha \cdot \boldsymbol{u}_i}{RT} + \frac{(\boldsymbol{e}_\alpha \cdot \boldsymbol{u}_i)^2}{2(RT)^2} - \frac{\boldsymbol{u}_i^2}{2RT}\right] \quad (2)$$

where $\alpha$ → 1-9; $w_\alpha$ are weights given by $w_1 = 4/9, w_{2-5} = 1/9,$ and $w_{6-9} = 1/36$; the discrete particle velocities $\boldsymbol{e}_\alpha$ by $\boldsymbol{e_1} = [0,0]$, $\boldsymbol{e_{2,4}} = [\pm c, 0]$, $\boldsymbol{e_{3,5}} = [0, \pm c]$, $\boldsymbol{e_{6,7}} = [\pm c, c]$ and $\boldsymbol{e_{8,9}} = [\pm c, -c]$; the lattice speed of sound by $c = \Delta x/\Delta t = \sqrt{3}c_s$; and, the speed of sound in the fluid medium by $c_s = \sqrt{RT} = 1/\sqrt{3}$. It is worth noting here that the LBE recovers the advection – diffusion equation for the ions in the hydrodynamic limit[37].

**Molecular speed dependent relaxation time.** Typically, in the BGK approximation the dimensionless relaxation time in the LBE is related to the equilibrium bulk kinematic viscosity $\nu_s^{eq}$ of the solvent or the ionic diffusion coefficients $D_i^{eq}$ as

$$\tau_s = \frac{\nu_s}{RT}\frac{\Delta t}{\Delta x^2} + 0.5 \quad \text{or} \quad \tau_i = \frac{D_i}{RT}\frac{\Delta t}{\Delta x^2} + 0.5 \qquad (3)$$

where $\nu_s$, $D_i$ → $\nu_s^{eq}$, $D_i^{eq}$, $\Delta x$ is the lattice spacing, $R$ the gas constant and T the temperature. Based on kinetic theory[46], in terms of the equilibrium mean free path $l_i^{eq}$ for the particles in the electrolyte mixture, the equilibrium ionic diffusion coefficients and solvent viscosity can be written as $D_i^{eq}, \nu_s^{eq} = (1/3)\bar{c}_i l_i^{eq}$ where $\bar{c}_i = \sqrt{8kT/\pi m_i}$ is the mean thermal speed of the particles of molecular mass $m_i$ at equilibrium, $T$ is the system temperature and $k$ is the Boltzmann constant. Away from equilibrium, the net macroscopic velocity $\boldsymbol{u}_i(\boldsymbol{r}, t)$ of the particles also represents a statistical average of the microscopic velocity of the fluid particles. As such, at any given time and position in the simulation domain the average local speed $c_i(\boldsymbol{r}, t)$ of the particles can be written as a sum of the molecular speed $\bar{c}_i$ due to random Brownian motion and the magnitude of the macroscopic velocity $\boldsymbol{u}_i(\boldsymbol{r}, t)$, i.e.

$$c_i(\boldsymbol{r}, t) = \bar{c}_i + u_i(\boldsymbol{r}, t) \qquad (4)$$

Thus, to compute the MSDRT for systems away from equilibrium, we define the local kinematic viscosity $\nu(\boldsymbol{r}, t)$ or the ionic diffusion coefficients $D_i(\boldsymbol{r}, t)$ as

$$\nu_s, D_i \equiv \frac{1}{3} c_i(\boldsymbol{r}, t) l(c_i) \tag{5}$$

where $l(c_i)$ is the local non-equilibrium mean free path. In the LB algorithm, since the local macroscopic velocity $\boldsymbol{u}_i(\boldsymbol{r}, t)$ resulting from the action of the external body force $\boldsymbol{F}_i$ on the $i^{th}$ type of particles is easily computed using equations for the macroscopic particle density $n_i$ and the momentum density

$$n_i = \sum_\alpha f_{i,\alpha} = \sum_\alpha f_{i,\alpha}^{eq} \tag{6}$$

$$n_i \boldsymbol{u_i} = \sum_\alpha \boldsymbol{e}_\alpha f_{i,\alpha} + \frac{\Delta t}{2} \boldsymbol{F}_i, \tag{7}$$

the problem of computing a MSDRT using equations (3)-(7) reduces to obtaining an estimate of the local non-equilibrium mean free path $l$ that depends on the local speed $c_i$ of the particles. This can be accomplished based on Tait's theory[44], with the underlying assumption that there are sufficient numbers of collisions in the system to allow for the existence of a local Maxwellian distribution of particle speeds. The number of collisions occurring in an interval of time $dt$ between pairs of molecules of type $i$ and $j$ with corresponding masses $m_i$ and $m_j$ in a mixture can be written as[43]

$$dt \xi_i \xi_j c_r d_{ij}^2 \cos\psi \sin\psi \, d\psi d\epsilon d\boldsymbol{c_i} d\boldsymbol{c_j} \tag{8}$$

such that $\boldsymbol{c_i}$, $\boldsymbol{c_j}$, $\psi$ and $\epsilon$ lie in the ranges $d\boldsymbol{c_i}$, $d\boldsymbol{c_j}$, $d\psi$ and $d\epsilon$ in the neighborhood of $\boldsymbol{c_i}$, $\boldsymbol{c_j}$, $\psi$ and $\epsilon$ respectively. Here, $\boldsymbol{c_i}$ and $\boldsymbol{c_j}$ are the particle velocities; $c_r$ is the magnitude of the relative velocity ($\boldsymbol{c}_{ji} = \boldsymbol{c}_j - \boldsymbol{c}_i$) of the particles; $\psi(= (\pi - \chi)/2)$ is the angle between the relative velocity $\boldsymbol{c}_{ji}$ and the unit vector $\boldsymbol{k}$ joining the centers of masses of the particles $i$ and $j$ at the point of their closest approach; $\chi$ is the angle of deflection between the relative velocity $\boldsymbol{c}_{ji}$ before collision and $\boldsymbol{c}_{ji}'$ after collision; $d_{ij}$ is the average diameter of the molecules of type $i$ and $j$; $\epsilon$ is the angle between the plane $LMN$ and the plane containing $AP'$ and the fixed z-axis $Oz$ (See Figure 1 for a description of the geometry of collision); and, $\xi_i$ and $\xi_j$ are the Maxwell-Boltzmann molecular velocity distribution functions given by

$$\xi_i = \eta_i \left(\frac{m_i}{2\pi kT}\right)^{3/2} exp\left(\frac{-m_i c_i^2}{2kT}\right) \tag{9}$$

where $\eta_i$ is the number density of the particles of type $i$. Now, the total number of collisions during $dt$ such that $\boldsymbol{c}_i$ lies in the velocity space volume element $d\boldsymbol{c}_i$ around $\boldsymbol{c}_i$ is proportional to $dt$ and to the number $\xi_i d\boldsymbol{c}_i$ of the particles of type $i$ which can be written as

$$\gamma_{ij}(c_i)\xi_i d\boldsymbol{c}_i dt \tag{10}$$

where $\gamma_{ij}(c_i)$ is the collision frequency for molecules of type $i$ moving at speed $c_i$ with molecules of type $j$. The collision frequency $\gamma_{ij}(c_i)$ is independent of the direction of $c_i$ and can be obtained by dividing equation (8) by $\xi_i d\boldsymbol{c}_i dt$ and integrating over all values of $\boldsymbol{c}_j$, $\psi$ and $\epsilon$ as[43]

$$\gamma_{ij}(c_i) = \eta_j(\boldsymbol{r}, t) d_{ij}^2 \sqrt{\frac{2\pi kT}{m_j}} \left[e^{-p_j{}^2} + \left(2p_j + \frac{1}{p_j}\right) Erf(p_j)\right] \tag{11}$$

where $p_j = c_i\sqrt{m_j/2kT}$, $Erf$ denotes the un-normalized error function, k the Boltzmann constant and $d_{ij}$ is the average molecular diameter of the molecules of type $i$ and $j$. Using equation (11) the mean free path of particles of type $i$ moving at speed $c_i$ in the fluid mixture can be calculated as $l_i(c_i) = c_i/\sum_j \gamma_{ij}$ which can then be substituted in equations (5) and (3) to compute the local non-equilibrium MSDRTs for use with the LBE. Since $\gamma_{ij}(c_i)$ and hence $l_i(c_i)$ are both independent of the direction of $\boldsymbol{c}_i$, the local non-equilibrium MSDRTs thus computed are isotropic.

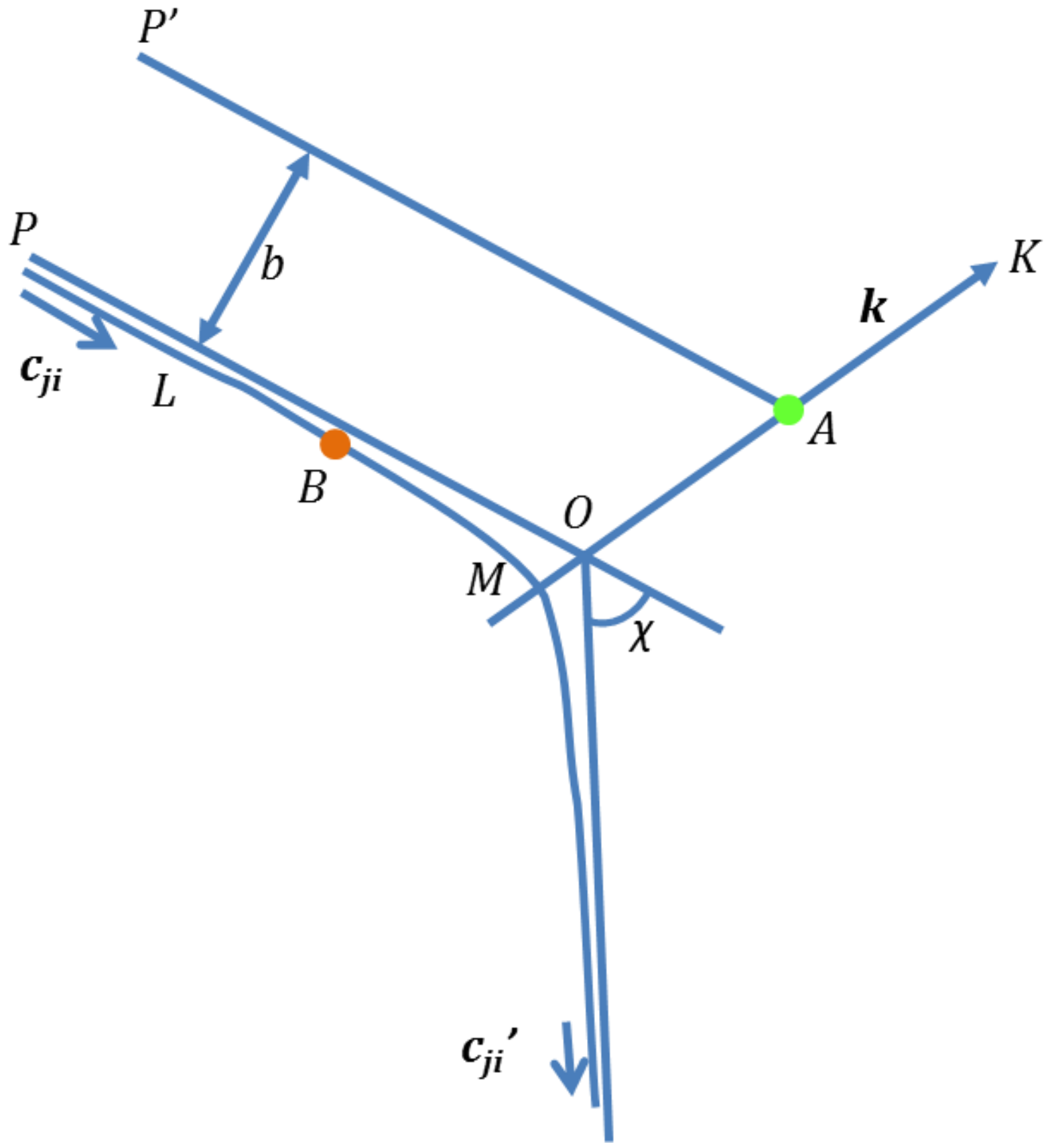

**Figure 1.** Geometry of a binary collision for particles of type $i$ (green circle) and $j$ (orange circle).

Based on the dependence of the MSDRT on the sum of the local mean thermal speed and the macroscopic speed of the particles, it may be mistakenly inferred that it would result in a violation of Galilean invariance. However, that is not true because the mean thermal speed depends only on the distribution of particle velocities irrespective of the inertial frame of reference employed and is therefore Galilean invariant[43] while the local macroscopic particle speed complies with Galilean invariance within an error on the order of the square of the Mach number $O(Ma^2)$ for the D2Q9 model employed here[47]. Therefore, the proposed LB formalism employing MSDRT is Galilean invariant for the small particle velocities observed in our simulation results.

**Entrance flow problem.** Free-slip boundary conditions were specified from the start of the simulation domain to the inlet of the channel for a distance of $100\ lu$ to simulate the effect of the reservoir while no-slip boundary conditions on the channel walls were specified for the rest of the simulation domain length i.e. $700\ lu$. The free-slip boundary conditions were implemented using specular reflection of the density distribution functions $f_{s,\alpha}$ such that there was no friction exerted on the fluid flow and the tangential component of the momentum remained unchanged[48]. The no-slip boundary conditions, designed to implement no tangential fluid flux along the channel walls, were implemented using bounce-back of the particle distribution functions[48]. Using non-equilibrium bounce-back of density distribution functions normal to the inlet and outlet boundaries[49], a velocity boundary condition of $\boldsymbol{u}_s = [0.05,0]\ lu$ was imposed at the free-slip inlet while at the outlet a constant pressure boundary condition was imposed by fixing the fluid density at $n_{s,out} = 1\ lu$. The dimensional mass of the fluid particles was specified as $m_s = 18.015\ amu$; temperature as $T = 298\ K$; density as $n_s = 1000\ kg/m^3$; and, diameter as $d_s = 1.92\ A$. The theoretical fully developed Poiseuille flow velocity profile was computed using $u_x = 4u_{\max}(y/h)(1 - y/h)$ where $h$ is the distance between the channel walls while the non-dimensional pressure was computed from the local fluid density using the equation of state for the fluid $p = \rho RT$ where $RT$ in the LB algorithm equals $1/3$.

**Electrolytic nanocapacitor.** The time evolution of the ionic density distribution functions in response to the applied electric potential on the nanocapacitor electrodes was simulated using the LBE (equation (1)) coupled to the nonlinear Poisson-Boltzmann equation. In our simulations, the electric field and potential, corresponding to the evolving ionic density distribution functions solved for using the LBE at each time step, were computed using the Lattice Poisson-Boltzmann Method (LPBM) proposed by Wang *et al*[36] for the solution of the nonlinear Poisson-Boltzmann equation. Our choice of LPBM for the solution of the nonlinear Poisson-Boltzmann equation was influenced by its ease of parallelization. However, other more efficient multigrid Poisson-Boltzmann solvers[50,51] may as well be employed equivalently. Dirichlet boundary conditions were applied for the electric potential $V_e$ on the electrode surfaces such that $V_e(x, y = 0) = +10\ mV$ and $V_e(x, y = h) = -10\ mV$ for all $t \geq 0$ while an initial condition of linearly varying electric potential from $+10\ mV$ at the cathode to $-10\ mV$ at the anode was specified throughout the simulation domain. To simulate perfectly blocking electrodes, no-flux boundary conditions were applied for the ions on the electrode surfaces using equation (7). On the simulation domain boundaries perpendicular to the electrode surfaces, periodic boundary conditions were implemented for distribution functions used in the LPBM for the computation of the electric potential and in the LBE for the simulation of ionic drift-diffusion. To account for the electric and viscous drag forces acting on the ions in the solvent, the discrete external force term $F_i$ in equation (1) was specified as[33,34]

$$F_i = \left(1 - \frac{1}{2\tau_i}\right)\left(\frac{\boldsymbol{e}_\alpha - \boldsymbol{u}_i}{RT} + \frac{\boldsymbol{e}_\alpha \cdot \boldsymbol{u}_i}{(RT)^2}\boldsymbol{e}_\alpha\right) \cdot [z_i e\boldsymbol{E} - 3\pi d_i \phi_i \mu_s(\boldsymbol{u}_i - \boldsymbol{u}_s)] f_{i,\alpha}{}^{eq} \tag{12}$$

where $\boldsymbol{E}$ is the electric field, $\mu_s$ the dynamic viscosity of the solvent and $\phi_i$ is the number fraction of the ions of the $i^{th}$ type in the electrolyte mixture. The first and the second terms in the square brackets in equation (12) correspond to the electric and the viscous drag forces acting on the ions respectively. For electrode spacings of $h = 50$ and $100\ nm$ at $t = 0$, ions in the electrolyte experienced an initial electric field of $E_y = 4 \times 10^5$ and $2 \times 10^5\ V/m$ respectively.

To understand the relaxation dynamics of the ions in the region between the two electrodes, simulations were carried out for various degrees of EDL overlap $\alpha(\equiv 10\lambda_D/h$, where $\lambda_D$ is the Debye length given by $\lambda_D = \sqrt{(\varepsilon kT/2n_i^b z_i^2 e^2)}$ and $n_i^b$ is the bulk ionic concentration), solvent viscosity $\mu_s$, electrode separations of $50$ and $100\,nm$ and cation to anion diffusion coefficient ratios of $1{:}\,1$ and $2{:}\,1$. The equilibrium anion diffusion coefficient was specified as $D_{an}^{eq} = 4 \times 10^{-9}\,m^2/s$ or $D_{an}^{eq} = 2 \times 10^{-9}\,m^2/s$ depending on the diffusion coefficient ratio while the equilibrium cation diffusion coefficient was fixed at $D_{cn}^{eq} = 4 \times 10^{-9}\,m^2/s$. Since the electrode spacing for almost all simulations was fixed at $h = 50\,nm$, the EDL overlap parameter was changed to $\alpha = 1.4,\ 0.9,\ 0.65, 0.2$ or $0.1$ by varying bulk electrolyte concentrations to $n^b = 1.89, 4.57, 8.76, 92.54$ or $370.19\,mM$ respectively. The effect of electrode spacing was studied by changing $h$ to $100\,nm$ while keeping the electrolyte concentrations fixed. The temperature for all simulations was fixed at $T = 298\,K$; solvent molecular mass $m_s$ and density $n_s$ were fixed at $18.015\,amu$ and $1000\,kg/m^3$ respectively; solvent relative dielectric constant was specified as $78.547$; molecular masses of hydrated ions were fixed at $113.065\,amu$ corresponding to five molecular masses of the solvent molecule in the hydration shell and an ionic mass of $m_{ion} = 22.99\,amu$. Corresponding to these conditions, the ionic diameters for the two anion to cation diffusion coefficient ratios of $1{:}\,1$ and $2{:}\,1$ were obtained from the Newton-Raphson method based iterative solution of the set of coupled quadratic equations described by equation (11) as $d_{cn} = d_{an} \approx 3.48\,A$ and $d_{cn} \approx 3.48\,A$; $d_{an} \approx 5.71\,A$ respectively. A lattice spacing of $\Delta x = 0.5\,A$ corresponding to a simulation time step of $\Delta t = 0.58\,ps$ was employed for all simulations.

## RESULTS AND DISCUSSION

**Entrance flow problem.** The entrance flow problem in isothermal hydrodynamics relates to the development of a fluid flow profile in a channel that connects two reservoirs with unequal hydrostatic pressures[45]. It is assumed that, as the fluid flows between the two reservoirs over time, there is no change in the pressure at either the inlet or the outlet of the channel. In the steady state, a parabolic flow velocity profile corresponding to Poiseuille flow develops in the channel. To validate the proposed MSDRT, the dynamics of a pure fluid flow were simulated in the absence of an external body force field using equation (1) until steady state was reached.

Simulation results in Figure 2a indicate the steady state pressure along the length of the channel. It can be seen that the hydrodynamic pressure close to the channel walls at the point of transition (at $100\ lattice\ units\ or\ lu$) from a free-slip to a no-slip boundary condition for the fluid flow is higher than it is at the center of the channel at the same position perpendicular to the channel wall. But, despite the pressure being higher closer to the point of transition, the fluid close to the walls would be forced to relax, if the BGK approximation is employed, at the same rate as the fluid at the center of the channel. A physically correct description of microscopic collision phenomena based on kinetic theory, however, must have a higher rate of relaxation for the fluid in a region of higher pressure on account of greater density than in a region of lower pressure. With the use of the MSDRT in the LBE, a comparison of the Figure 2a-c clearly shows that, as expected, the non-dimensional relaxation frequency $\omega_s(r, t)$ is higher in regions of higher pressure and lower speed and $vice\ versa$. Finally, Figure 2d shows that the transverse flow speed profile across the width of the channel at the outlet exactly matches the theoretically obtained parabolic profile for the plane Poiseuille flow.

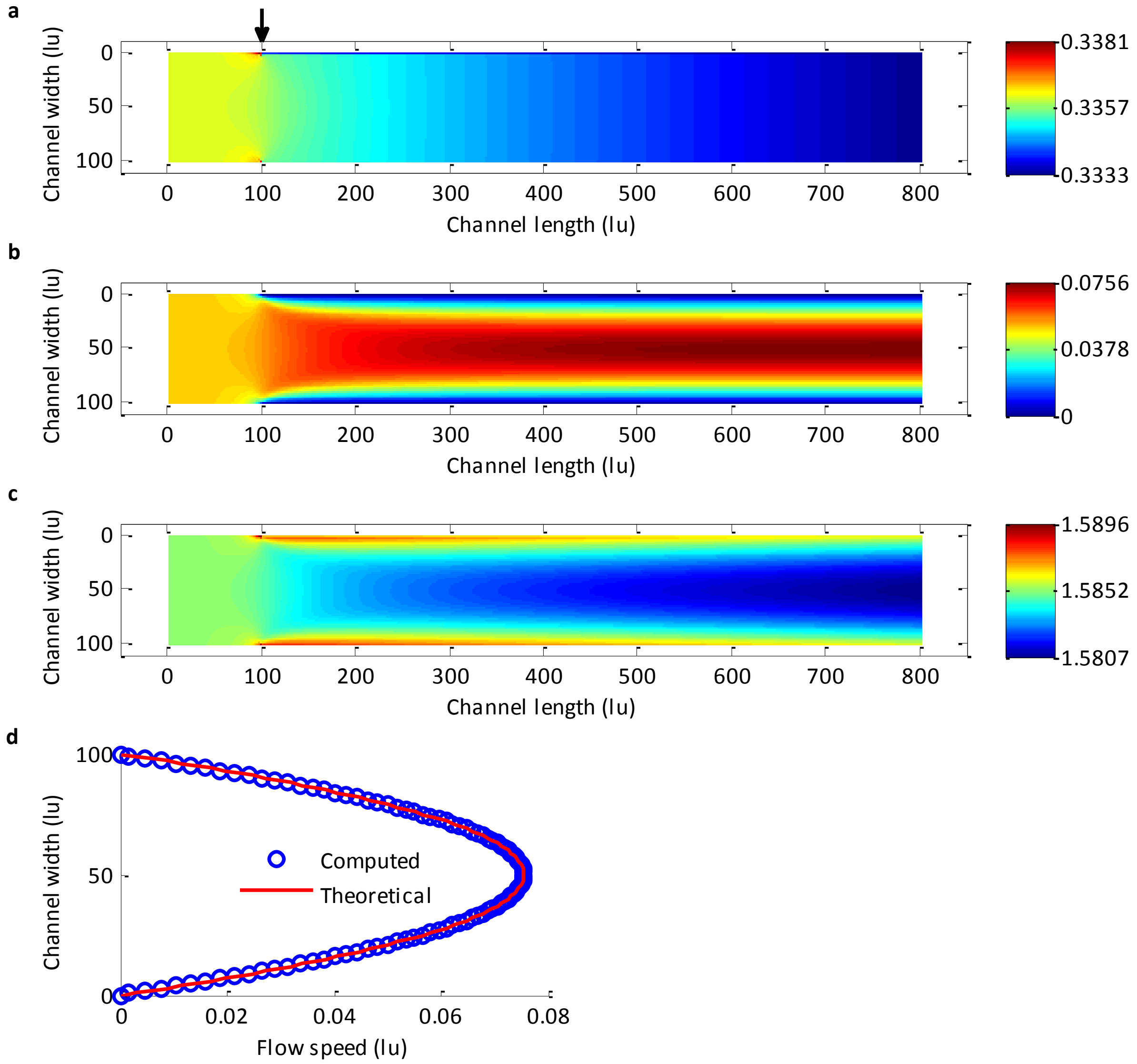

**Figure 2.** Entrance flow in a channel. Non-dimensional (in lattice units, $lu$) (a) Pressure, (b) Flow speed, (c) Collision frequency, and, (d) Poiseuille flow velocity profile across the channel width at the outlet in the channel. The vertical arrow in Figure 2(a) marks the point of transition from free-slip boundary condition on the left to no-slip boundary condition on the channel walls on the right.

**Electrolytic nanocapacitor.** For the simulation of the charging dynamics of the electrolytic nanocapacitor, a primitive model of a symmetric $1:1$ electrolyte was considered. The ions in the solvent were described as point particles and the effect of the solvent in the simulations was included by describing it as a background medium with a characteristic dielectric constant, a constant density and viscosity. Although the LB algorithm presented here can be easily extended to simulate the coupled ion-solvent relaxation dynamics as well, a deliberate choice was made to exclude the dynamics of the solvent from the simulations for the sake of simplicity and as such the solvent velocity, $\boldsymbol{u}_s$, was approximated to zero. However, for the purpose of computing the MSDRTs for the ionic distribution functions $f_{i,\alpha}$, hydrated ions with an effective mass $m_i$ and finite diameter $d_i$ were considered for use in equation (11). The time evolution of ionic density distributions, in the overlapping EDLs of the electrolytic nanocapacitor in response to an applied electric potential step at $t = 0$, was simulated using the discrete LBE described in equation (1). The resulting electric force, acting on the uniformly distributed electrolyte ions at $t = 0$, caused them to accelerate towards electrodes of opposite polarities and impinge on the capacitor electrodes. The subsequent ion relaxation dynamics in the region between the two electrodes were then investigated.

*Space-averaged current density.* The effect of EDL overlap $\alpha(\equiv 10\lambda_D/h$, where $\lambda_D$ is the Debye length and $h$ is the separation between the two electrodes of the capacitor) and solvent viscosity on the space-averaged current density

$$J_{avg}(t) = \frac{1}{h}\int_0^h dy \sum_i z_i e n_i(\boldsymbol{r},t) u_i(\boldsymbol{r},t)$$

is depicted in Figure 3a-c. It is observed that, as the degree of EDL overlap $\alpha$ is reduced from 1.4 to 0.65 and then to 0.1, the behavior of the space-averaged current density changes gradually from oscillatory to monotonic accompanied by a decrease in the maximum amplitude. The effect of increase in solvent viscosity $\mu_s$ in going from $0.0008$ to $0.0018\, Pa.s$ is to dampen the oscillatory behavior for $\alpha = 1.4$, $0.65$ and reduce the amplitude of the oscillations (Figure 3a,b) while it increases the relaxation time to equilibrium for the space-averaged current density in the case of $\alpha = 0.1$ (Figure 3c). The oscillatory behavior in the case of $\alpha = 1.4, 0.65$ is seen to persist for $\mu_s = 0.0008, 0.0013\, Pa.s$ much longer than what might be expected (fig 2a,b). Figure 3d,e shows that increasing the electrode separation from $h = 50$ to $100\, nm$ decreased the maximum amplitude of $J_{avg}$ for both bulk ionic concentrations of $n_b = 4.6$ and $92.5\, mM$. Doubling the electrode separation also increased the period of oscillations in $J_{avg}$ by a factor of ~1.5-2.0 for $n_b = 4.6\, mM$ and doubled the relaxation time to equilibrium for $n_b = 92.5\, mM$. The effect of the ratio of equilibrium cation to anion diffusion coefficients on $J_{avg}$ for a solvent viscosity of $0.0018\, Pa.s$ is shown in Figure 3f,g. A reduction of anion diffusivity by half reduced peak $J_{avg}$ for both $\alpha = 1.4, 0.1$ and was accompanied by a slight increase in the relaxation time for $\alpha = 0.1$ (Figure 3f,g). The disparity in diffusivity of ions also introduced very small amplitude persistent oscillations in $J_{avg}$ for $\alpha = 1.4$ compared to when the ionic diffusivities were equal.

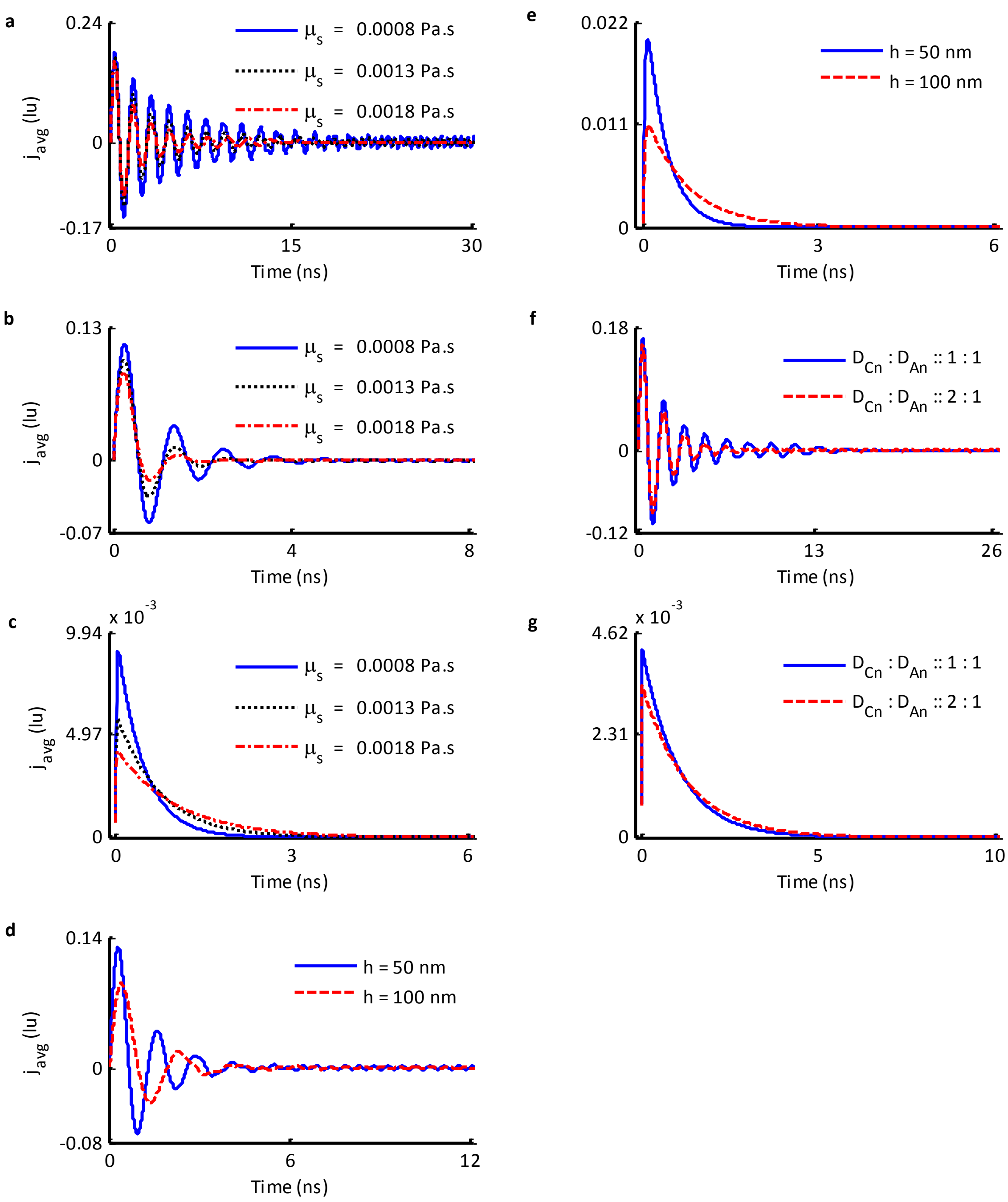

**Figure 3.** Space-averaged current density as a function of time. Effect of EDL overlap and solvent viscosity with $h = 50\,nm$ and $D_{Cn}: D_{An}$ :: 1: 1, (a) α = 1.40, (b) α = 0.65, (c) α = 0.10; Electrode separation with $\mu_s = 0.0013\,Pa.s$ and $D_{Cn}: D_{An}$ :: 1: 1, (d) $n_b = 4.6$ mM, (e)

$n_b = 92.5$ mM; Ratio of cation to anion diffusion coefficient with $\mu_s = 0.0018\, Pa.s$ and $h = 50\, nm$, (f) α = 1.4, (g) α = 0.10.

In terms of underlying physics, the decrease in peak amplitude of $J_{avg}$ with an increase in solvent viscosity $\mu_s$ (Figure 3a-c) and a halving of the anion diffusion coefficient $D_{an}$ (Figure 3f,g) are similar and result from a reduction in the mobility of ions as both viscosity and diffusivity are empirically related to each other and mobility through the Nernst-Einstein and Stokes-Einstein relations $D_i = \mu_i kT/z_i e = kT/3\pi d_i \mu_s$. A decrease in ionic mobility, therefore, leads to a decreased conductivity of the electrolytic medium between the capacitor electrodes resulting in an increase in the potential drop across the electrolyte and a consequent decrease in the peak amplitude of $J_{avg}$. This also explains faster relaxation to equilibrium for the monotonic case of $\alpha = 0.1$ with decreasing viscosity of the solvent as more mobile ions charge up the electrodes faster (Figure 3c). For $\alpha = 1.4$ and $0.65$, a decrease in viscosity and the consequent increase in ionic mobility causes the ions to oscillate longer (Figure 3a,b).

Now, for a monotonically charging capacitor the current density as a function of time is $J = J_0 e^{-t/\tau_c}$ where $\tau_c = RC = (h/\sigma A)C$ is the characteristic relaxation time for the capacitor, $A$ the effective area of the capacitor electrodes, $C$ the capacitance, $\sigma$ the conductivity of the dielectric medium between the two electrodes and $J_0$ the peak current amplitude. Thus, a doubling of the electrode spacing $h$ must result in a doubling of the characteristic charging time $\tau_c$ as well (Figure 3e). Also, an increase in the resistance $R = (h/\sigma A)$ accounts for the decrease in the maximum value for $J_{avg}$ (Figure 3d,e). Similarly, doubling the distance between the capacitor electrodes doubles the time taken by ions in moving from one electrode to another thereby resulting in a near doubling of the period of oscillations for $J_{avg}$ (Figure 3d). The results for $J_{avg}$ indicate that in the limit of thin EDLs the monotonic charging dynamics for an electrolytic nanocapacitor as in macroscopic electrochemical systems is recovered while for large EDL overlaps an oscillatory behavior is observed.

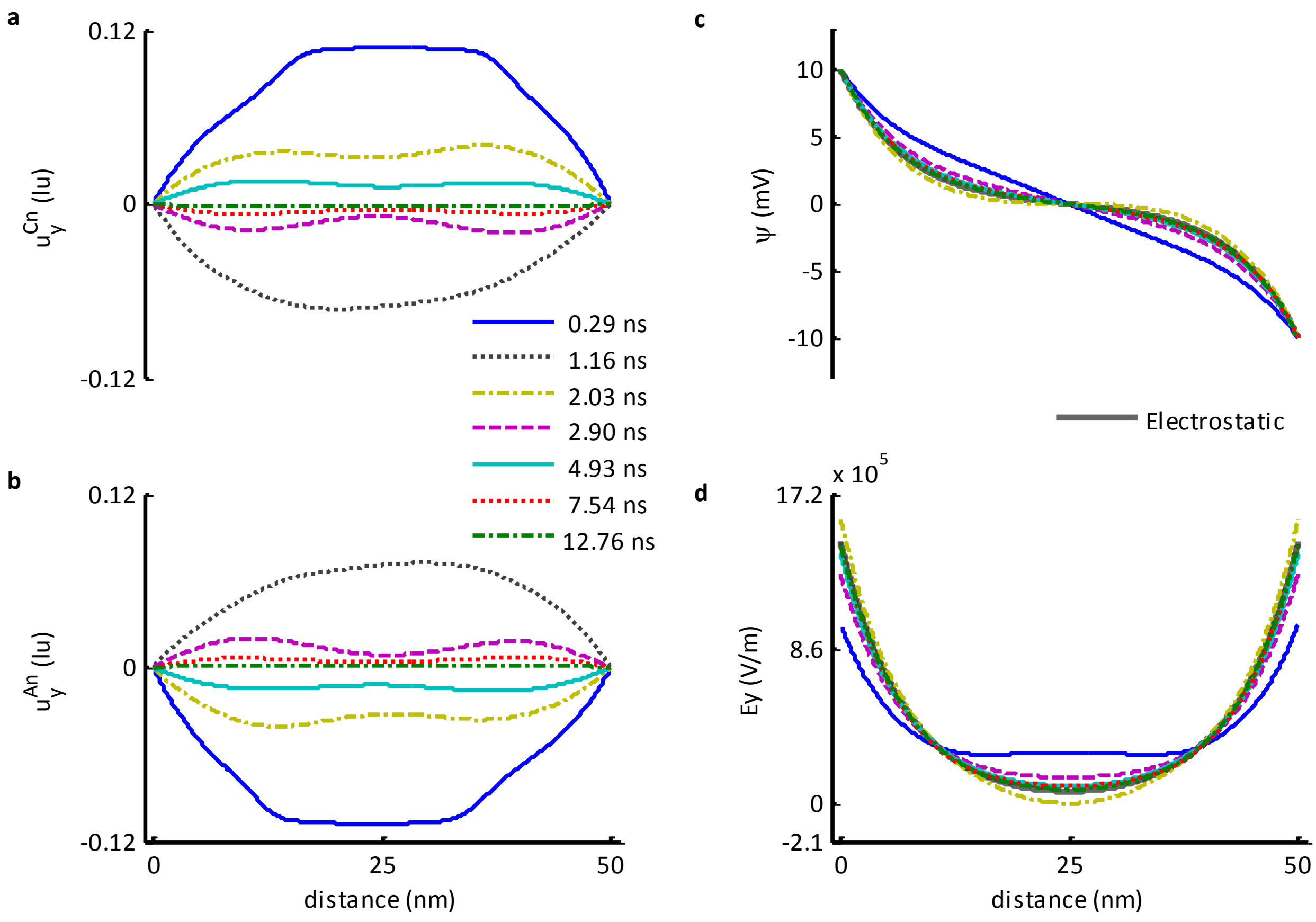

**Figure 4.** Oscillatory behavior of ions for large EDL overlap (Also, see Movie S1). (a) Cation velocities, (b) Anion velocities, (c) Electric potential, and, (d) Electric field as a function of time for $\alpha = 1.4$, $\mu_s = 0.0018\ Pa.s$, $h = 50\ nm$, and $D_{Cn}: D_{An} :: 1:1$.

*Collective oscillatory dynamics of ions.* For an EDL overlap of $\alpha = 1.4$, Figure 4a,b depicts the damped harmonic motion of ions as the ionic velocities periodically reverse direction between the perfectly blocking electrodes and then decay to zero over time. As a consequence, the electric potential and field profiles also evolve in time in an oscillatory fashion towards their respective equilibrium electrostatic profiles (Figure 4c,d and Supporting Information Movie S1). In the case of $\alpha = 0.1$, Figure 5a-d and Supporting Information Movie S2 show a monotonic convergence to equilibrium electrostatic values for these variables. The force that ions experience at any point between the two electrodes of the capacitor is proportional to the electric field $\boldsymbol{E}$ seen by the ions at that location. A comparison of the electric field profiles in the gap between the capacitor electrodes in Figures 3d and 4d for the two cases of EDL overlap shows that both exhibit a trough in the middle. This trough in the case of an EDL overlap of $\alpha = 1.4$ is almost parabolic in shape with a non-zero electric field in the center while the one for $\alpha = 0.1$ is flat bottomed with a field that vanishes at the center of the capacitor. So, in terms of an analogy, the oscillatory and monotonic motions of ions can be likened to the motion of a ball rolled in parabolic and flat bottomed troughs with frictional surfaces under the influence of gravity from the top edges of the troughs respectively. In the parabolic case, the ball executes a to-and-fro oscillatory motion before coming to rest while in the flat-bottomed case it comes to rest more or less monotonically. Thus, this explains that it is the parabolic or near parabolic shape of the electric field profile in conjunction with a non-vanishing electric field, in the case of overlapping EDLs, at the center of the nanocapacitor that is responsible for the observed oscillatory behavior.

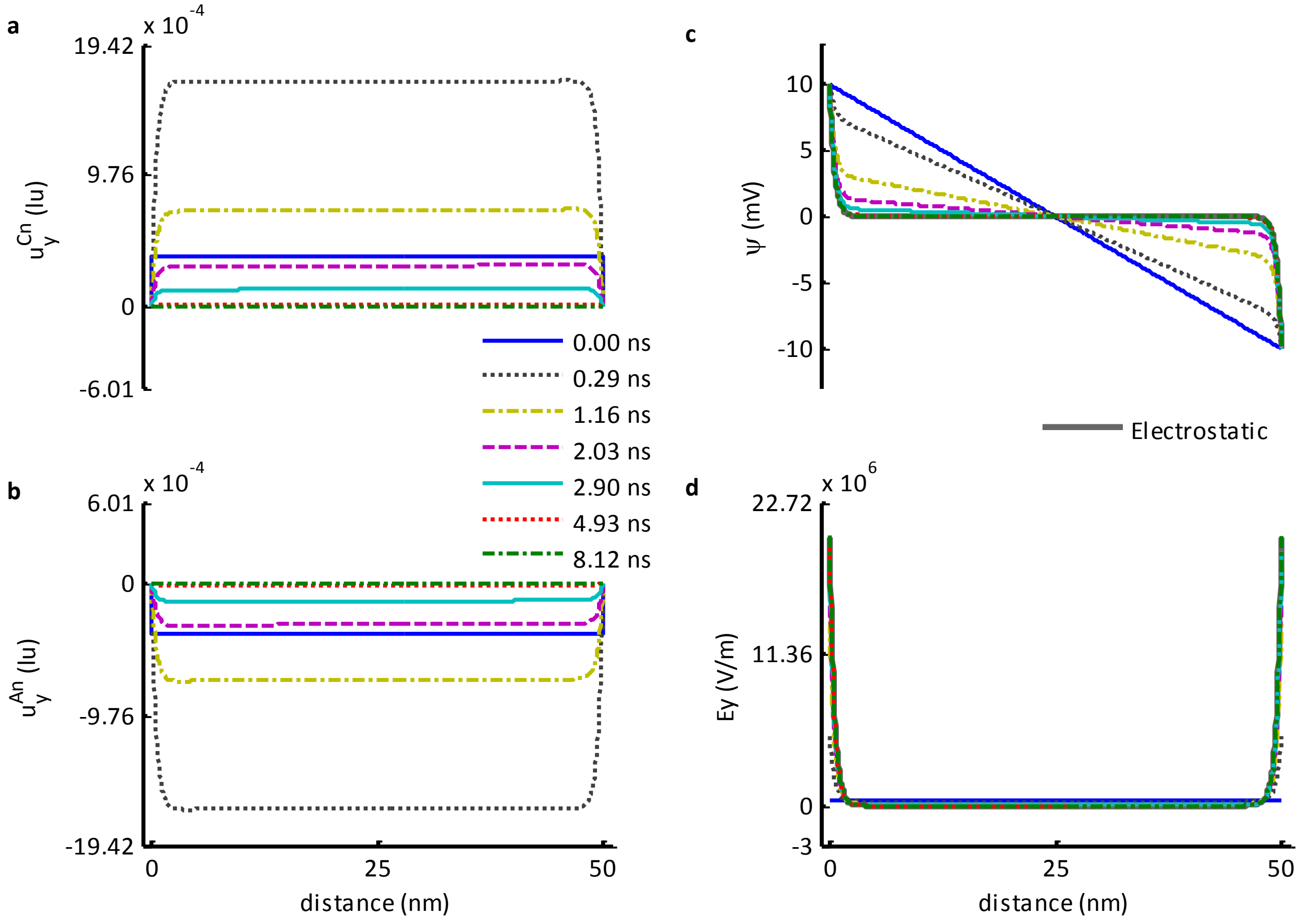

**Figure 5.** Monotonic charge relaxation for small EDL overlap (Also, see Movie S2). (a) Cation velocities, (b) Anion velocities, (c) Electric potential, and, (d) Electric field as a function of time for α = 0.10, $\mu_s = 0.0018\, Pa.s$, $h = 50\, nm$, and $D_{Cn}: D_{An} :: 1:1$.

*Anomalous spatial oscillations of electric field.* A persistence of oscillations in $J_{avg}$ was observed in three cases for an EDL overlap of $\alpha = 1.4$: (i) for $\mu_s = 0.0008\, Pa.s$, (ii) $\mu_s = 0.0013\, Pa.s$ (Figure 3a), and, (iii) for $\mu_s = 0.0018\, Pa.s$ when an asymmetry was introduced by changing the cation to anion diffusion coefficient ratio from $1:1$ to $2:1$ (Figure 3f). Since the origin of persistent oscillations is similar for cases (i) and (ii), besides case (iii), only the case pertaining to $\mu_s = 0.0008\, Pa.s$ is analyzed further.

Considering case (i) first, the ionic velocities plotted in Figure 6a,b show a general oscillatory behavior for a period up to 11 ns which is reflected in the dimensionless collision frequencies for the ions, electric field, the spatial cross-correlation ($J_y^{corr}$) of ionic flux densities ($J_{Cn}^f$ and $J_{An}^f$) and space-averaged current densities ($J_{avg}$, $J_{Cn}^{avg}$ and $J_{An}^{avg}$) (Figure 6a-e and Supporting Information Movie S3). Thereafter, between 11 and 20 ns there spontaneously develop two nodes at roughly 1/3rd of the electrode spacing from either electrode where the ionic velocities go to zero and switch signs (Figure 6a,b). The region of these two nodes between the nanocapacitor electrodes is marked by intense collisions as evidenced by the presence of two sharp peaks at roughly 1/3rd of the separation between the two electrodes from each electrode (Supporting Information Movie S3). Once the persistent oscillations are fully developed, between 58 and 93 ns, both cations and anions self-organize to result in their velocities forming a single node between the capacitor plates. At these nodes the ionic velocities switch sign at any one given instance of time (Figure 7a-d and Movie S3) and the position of this node in the center of the nanocapacitor is again marked by intense collisions between ions (Figure 7a,b). On either side of this node there exist regions where the collision frequencies are much lower and the ionic velocities are high. In these regions of low collisions the ionic velocities switch sign at different instances of time. Next, upon examination of the spatial cross-correlation $J_y^{corr}$of the y-component of the ionic flux densities $J_{Cn}^f(t)$ and $J_{An}^f(t)$ (Figure 6d) in the time regime (0-11 ns) when $J_{Cn}^{avg}$ and $J_{An}^{avg}$ are in-phase (Figure 6e), one observes that spatial cross-correlation $J_y^{corr}$ stays negative and the electric field amplitude for $E_y$ exhibits oscillatory behavior about the equilibrium electrostatic electric field in the center of the simulation domain (Figure 6c). The negative values of $J_y^{corr}$ in Figure 6d are consistent with the in-phase behavior of $J_{Cn}^{avg}$ and $J_{An}^{avg}$ because for a given electric field at a point oppositely charged ions must move in opposite directions and hence the negative correlation in their respective fluxes. Between 11 and 20 ns, however a positive correlation between the ionic flux densities $J_{Cn}^{f}(t)$ and $J_{An}^{f}(t)$ begins to manifest itself (Movie S3 and Figure 6d). For the spatial cross-correlation $J_y^{corr}$ in the latter out-of-phase time regime (58-93 ns) for $J_{Cn}^{avg}$ and $J_{An}^{avg}$, positive values are observed with a peak at a spatial lag or separation of about 25.3 nm (Figure 7d). Also, the electric field $E_y$ starts to exhibit spatial oscillations in this time regime (Figure 7c and Movie S3) unlike the case of simple oscillations in amplitude at the center of the nanocapacitor when $J_{Cn}^{avg}$ and $J_{An}^{avg}$ are in-phase (Figure 6c). It is seen that the minima of these spatial oscillations are again separated roughly by 25.3 nm.

Similar positive spatial cross-correlation of ionic flux densities, although much weaker than in case (i), is also observed for case (iii) with an EDL overlap of $\alpha = 1.4$, cation to anion diffusion coefficient ratio of $2:1$ and a solvent viscosity of $\mu_s = 0.0018\, Pa.s$ (Supporting Information Movie S4). Since such a behavior is absent in the case of $1:1$ diffusion coefficient ratio under similar conditions, the effect of introduction of an asymmetry in the diffusion coefficients or mobility of the ions is to cause persistent spatial oscillations in the electric field $E_y$ and a positive correlation in the ionic flux densities $J_{Cn}^{f}$ and $J_{An}^{f}$. Compared to the three cases discussed above, an exact contrast is presented in the case of an EDL overlap of $\alpha = 0.1$, cation to anion diffusion coefficient ratio of $1:1$ and a solvent viscosity of $\mu_s = 0.0018\, Pa.s$ (Movie S2) where the linear negative $J_y^{corr}$ monotonically goes to zero while the system marches to equilibrium as evidenced in the monotonic time evolution of the electric field $E_y$.

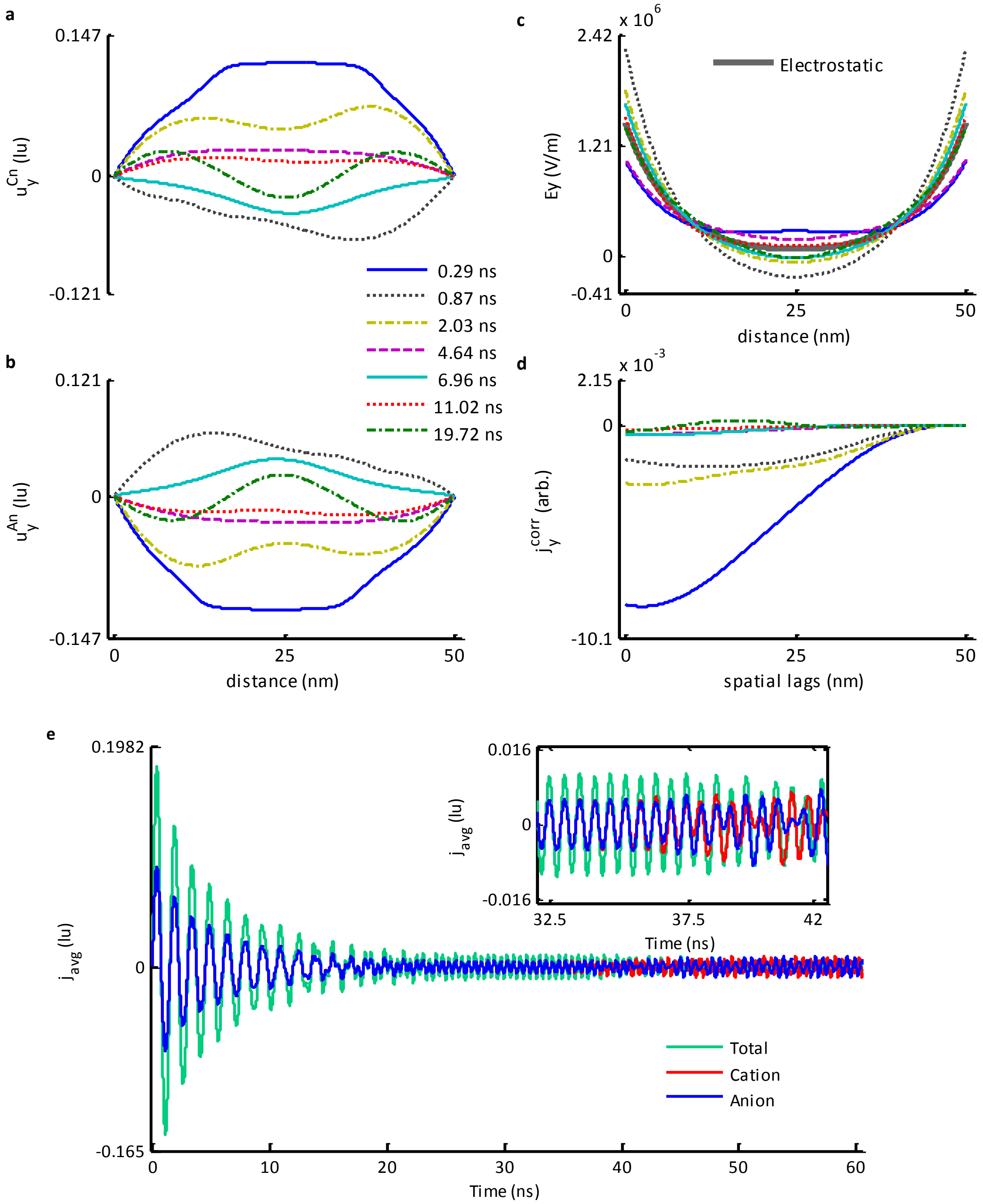

**Figure 6.** Oscillatory charge relaxation before the onset of plasma-like collective behavior (Also, see Movie S3). (a) Cation velocities, (b) Anion velocities, (c) Electric field, (d) Spatial cross-

correlation of cation and anion flux densities. (e) Space-averaged total, cation and anion current densities as a function of time for α = 1.40, $\mu_s = 0.0008\ Pa.s$, $h = 50\ nm$, and $D_{Cn}: D_{An} :: 1:1$ (Inset) Transition of space-averaged cation and anion current densities from in-phase to out-of-phase behavior marking the onset of plasma-like collective behavior.

Thus, it becomes apparent that when $J_{Cn}^{avg}$ and $J_{An}^{avg}$ are out-of-phase, the oppositely charged ions in the two opposite halves of the simulation domain exhibit positively correlated motions and form a sort of dynamic standing wave-pattern (Movie S3). It is this collective behavior that results in plasma like spatial oscillations in the electric field. But then, what is the reason behind such an anomalous collective behavior? The reason becomes clear when one analyzes the behavior of a dielectric slab upon insertion into a capacitor (Figure 7e). A dielectric slab, when slowly introduced into a capacitor, experiences a non-uniform electric field that exerts a force $\boldsymbol{F_d} = \boldsymbol{\nabla}(\boldsymbol{p_d}.\boldsymbol{E})$ on it that tends to drive it into a region of higher electric field. Since the field inside a dielectric slab points in a direction opposite to that of the electric field between the capacitor plates, the insertion of the dielectric slab has the effect of lowering the net electric field in the region occupied by it. Now consider this dielectric slab to be made up of oppositely charged ions interacting with each other to form temporary dipoles. With this, if the binding energy of the dipoles in the dielectric slab were less than the interaction energy of the individual ions (forming the dipole) with the external electric field of the capacitor then the dipoles will be pulled apart into individual ions and forced to move away (in the same direction) from the region of higher electric field as both types of ions are in a position that is far away from the equilibrium for the individual ions.

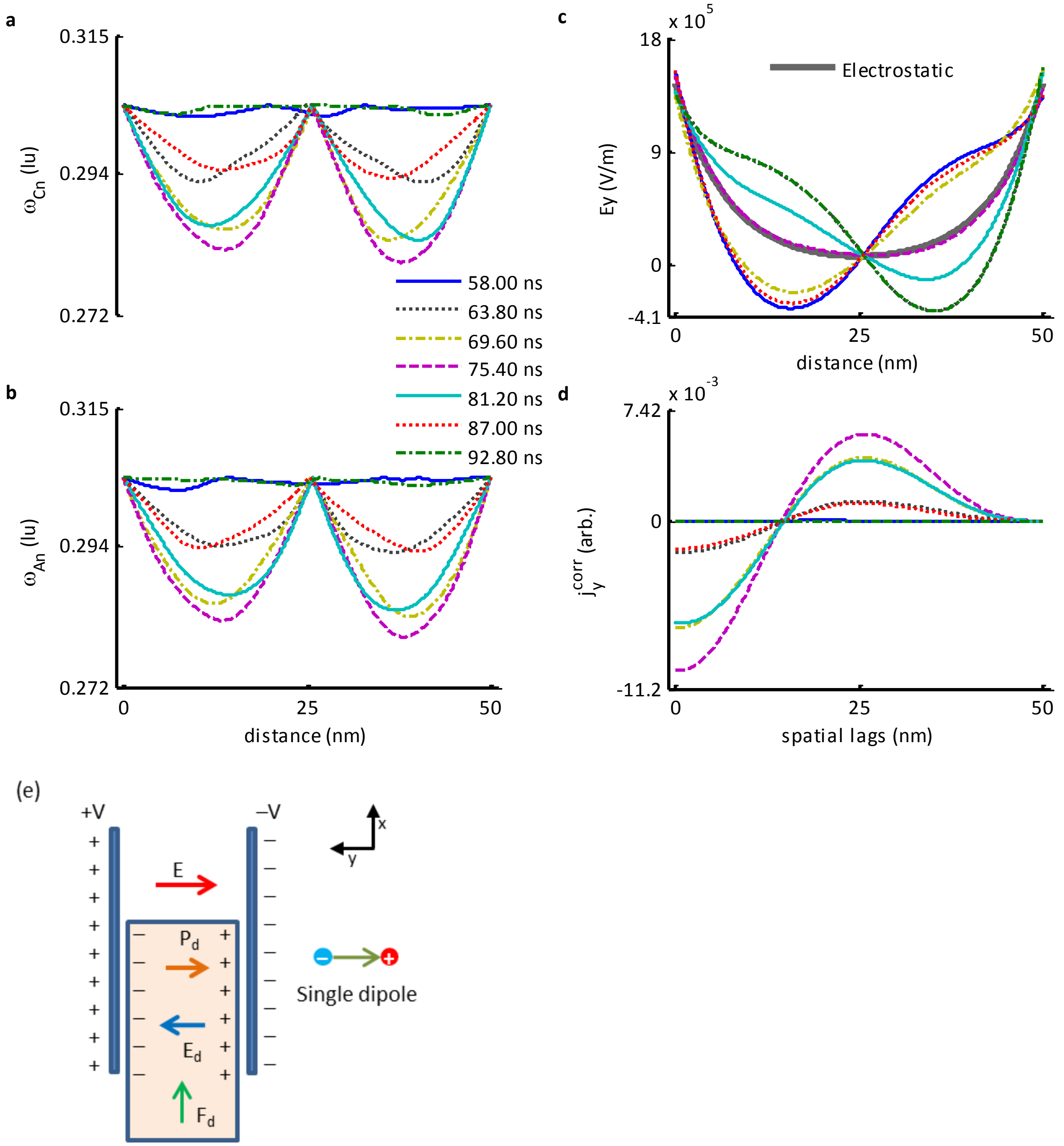

**Figure 7.** Anomalous oscillations in electric field after the onset of plasma-like collective behavior (Also, see Movie S3). (a) Cation collision frequencies, (b) Anion collision frequencies, (c) Electric field, and, (d) Spatial cross-correlation of cation and anion flux densities as a function of time for α = 1.40, $\mu_s = 0.0008\, Pa.s$, $h = 50\, nm$, and $D_{Cn}{:}D_{An} :: 1{:}1$. (e) Dielectric slab in the non-uniform electric field of a parallel plate capacitor.

Consider next a set of conditions with a region of lower electric field and low ionic velocities accompanied by intense collisions in the center of the nanocapacitor that allows them to interact again to form dipoles. Now, if the field between the capacitor plates be non-uniform as is the case for the electrolytic nanocapacitor, the dipolar interaction between the ions would again force them to move in to a region of higher electric field but this time in the opposite direction because of a residual momentum from their previous interaction with a region of high electric field. The net electric field in the region occupied by the ions interacting like dipoles would also be lowered until the time they again encounter a region of high electric field sufficiently strong to break them apart and cause a repeat of the cycle described above. Such a repetitive behavior would thus result in an oscillatory behavior of the electric field in space as a consequence of the spatially correlated motion of oppositely charged ions the kind of which was observed in Figure 7a-d. Thus, it can be concluded that the persistent spatial oscillations observed in $E_y$ and $J_{cn}^{avg}$ and $J_{an}^{avg}$ are most likely due to a dipolar interaction occurring between the ions that results in a plasma like collective behavior.

The plasma frequency $\omega_p$ for an electrolyte in terms of the Debye length $\lambda_D$ is given by $\omega_p = \lambda_D^{-1} \sqrt{(kT/m^*)}$ where $m^*$ is the harmonic mean of the effective cation and anion masses[39]. An EDL overlap of $\alpha = 1.4$ corresponding to an ionic concentration of $n_b = 1.89\ mM$ gives a plasma frequency of $\omega_p \cong 20.6\ GHz$ while the collision frequency for the ions in a solvent of density $n_s = 1000\ kg/m^3$ is of the order of $\nu_c \approx 5\ THz$. Since for such an electrolyte the collision frequency $\nu_c \gg \omega_p$, the current density is always in phase with the applied electric field and a plasma oscillation cannot be excited[52]. However, even with an oscillation frequency of $1.85\ GHz$ $(\sim\omega_p/10)$ for the in-phase ionic current densities $J_{cn}^{avg}$ and $J_{an}^{avg}$ just before the onset of the out-of-phase behavior, plasma like spatial oscillations are observed (Figure 6e inset). Such a behavior is clearly anomalous and therefore the results pertaining to plasma-like collective behavior presented here for overlapping EDLs in an electrolytic nanocapacitor appear to be purely an effect of nanoscale confinement.

## CONCLUSIONS

The use of a single relaxation time under BGK approximation in the LBE greatly simplifies computation. However, in general, the use of single relaxation time in the BGK approximation in LBM based simulations is known to suffer from problems of numerical instability and spurious artifacts especially in systems away from equilibrium that involve strong spatial variations in body forces, particle densities or temperature[4,53-57]. Various attempts at solution of these problems have been made using multiple relaxation time (MRT)[58], two relaxation time (TRT)[59] and the entropic[60] LB models. These approaches either employ constant relaxation times based on bulk transport coefficients or a variable over-relaxation parameter derived from them using entropic constraints. Since relaxation times in MRT and TRT LB models are based on constant bulk transport coefficients, none of these two methods aimed at improving the numerical stability of LBM simulations recover a correct microscopic description of the collision phenomena, and, besides being computationally expensive, it is not clear if entropic LB models do that either. Other variants of the LB algorithm have explored heuristically introduced density dependence into the relaxation time through a simple division of the constant BGK relaxation time by the position dependent local fluid density[61] and constant relaxation time approaches based on tunable diffusivities for binary mixtures through a consideration of mutual and cross-collisions separately that are aimed at recovering prescribed transport properties of a binary mixture[62-64].

The MSDRT proposed here, however, allows for continuously varying relaxation times in the simulation domain that recovers a correct microscopic description of the collision phenomena while simultaneously retaining the simplicity of the LB algorithm in the BGK approximation. Also, since it incorporates the local macroscopic speed of the particles, their number densities, molecular masses and the temperature in the computation of the relaxation time for the LBE, it may be better suited for the simulation of the dynamics of complex systems such as interfaces in multiphase fluids[53,54], mesoscale electrochemical or electrokinetic systems[6,7] with overlapping EDLs[4,55] that suffer from problems of numerical instabilities or spurious artifacts and involve strong spatial variations in body forces[4,55], particle densities[53,54,57] or temperature[56]. The effectiveness of the MSDRT proposed here is demonstrated through the transient simulations for the highly nonlinear charge relaxation dynamics of an electrolytic nanocapacitor with spatially non-uniform force fields generated due to inhomogeneous charge distributions in overlapping EDLs.

Additionally, we have successfully shown that the proposed LB method, using the continuously varying MSDRT, can be employed to simulate and explore the fundamental physics of non-equilibrium charge relaxation dynamics in mesoscale electrochemical systems wherein Nernst-Einstein relation is known to be violated. So far, this had been one of the most challenging unsolved problems in non-equilibrium electrokinetics. Further, our results predict the presence of hitherto unobserved phenomena of spatial oscillations in electric field arising from a spontaneous dipolar interaction of ions in response to a step potential applied across an electrolytic nanocapacitor under conditions of large EDL overlap and low solvent viscosity.

It is for these aforementioned reasons we believe that the LB method presented here will open up avenues for advancing an understanding of the phase transformation, transport and interface phenomena in electrochemical, electrokinetic and microfluidic systems critical to the development of technologically advanced mesoscale devices and synthesis of novel nanomaterials.

*Author contributions.* VT was responsible for designing the study, implementing the parallel codes for LBM, analysis and interpretation of the simulation results and preparation of the manuscript under the guidance and supervision of JJH.

*Conflict of interest:* The authors declare no competing financial interests.

*Acknowledgment.* The authors gratefully acknowledge funding and support from NIH grants R01NS050452, R01EB005459 and UH2TR000516, XSEDE start-up allocation through NSF OCI-1053575 and STOKES ARCC at the University of Central Florida.

*Supporting information available:* Movie S1: Oscillatory charge relaxation dynamics of an electrolytic nanocapacitor for large EDL overlap; Movie S2: Monotonic charge relaxation dynamics of an electrolytic nanocapacitor for small EDL overlap; Movie S3: Anomalous plasma like collective behavior at large EDL overlap and low solvent viscosity; Movie S4: Induced plasma like oscillations in charge relaxation dynamics due to an asymmetry in ionic diffusion coefficients. This material is available free of charge *via* the Internet at http://pubs.acs.org.

TABLE OF CONTENTS GRAPHIC

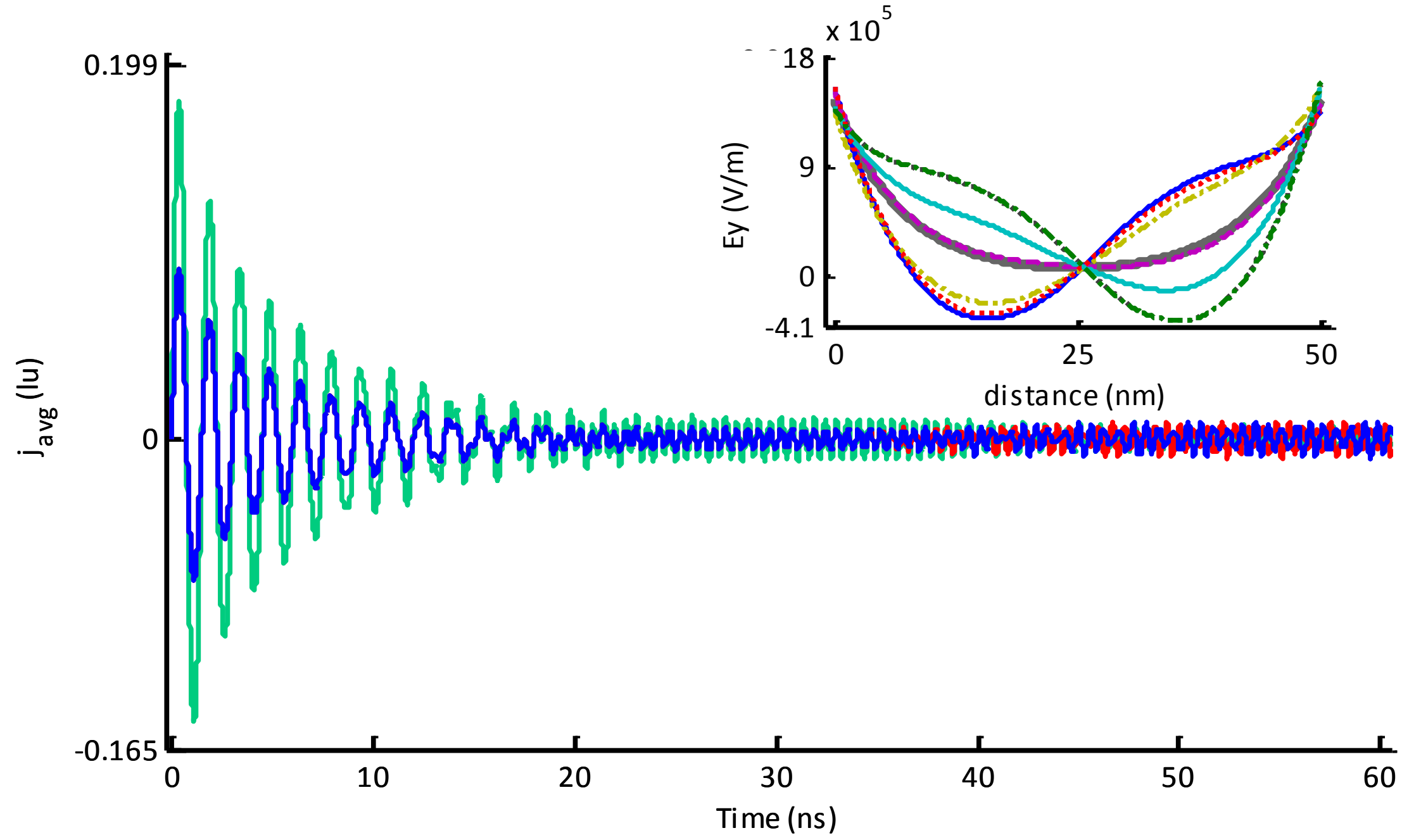